\documentclass[nofootinbib,twocolumn]{revtex4-1}

\usepackage[pdfencoding=auto, psdextra]{hyperref}
\usepackage[utf8]{inputenc}
\usepackage{amsmath,amssymb,amsthm}

\usepackage{cleveref}

\usepackage{scalerel,stackengine}
\stackMath
\newcommand\reallywidehat[1]{%
\savestack{\tmpbox}{\stretchto{%
  \scaleto{%
    \scalerel*[\widthof{\ensuremath{#1}}]{\kern-.6pt\bigwedge\kern-.6pt}%
    {\rule[-\textheight/2]{1ex}{\textheight}}
  }{\textheight}%
}{0.5ex}}%
\stackon[1pt]{#1}{\tmpbox}%
}

\DeclareMathOperator{\sgn}{\mathrm{sgn}\,}

\def\R{\mathbb{R}}

\def\fom{\mathring{\omega}}
\def\fe{\mathring{e}}
\def\fq{\mathring{q}}

\def\tm{\tilde{\mu}}
\def\hh{\hat{H}}

\newcommand{\SU}{\mathrm{SU(2)}}

\newcommand{\hc}{\text{h.c.}} 
\newcommand{\cc}{\text{c.c.}} 

\newcommand{\BI}{\gamma}
\newcommand{\BO}{\mathcal{O}} 
\newcommand{\lo}{o} 
\newcommand{\lp}{\ell_p} 

\def\beq{\begin{equation}}
\def\eeq{\end{equation}}
\def\bq{\begin{equation*}}
\def\eq{\end{equation*}}
\def\bs{\begin{split}}
\def\es{\end{split}}

\begin{document}


\title{Uniqueness of minimal loop quantum cosmology dynamics}
\author{Jonathan Engle}
\email[]{jonathan.engle@fau.edu}
\author{Ilya Vilensky}
\email[]{ilya.vilensky@fau.edu}
\affiliation{Florida Atlantic University, 777 Glades Road, Boca Raton, FL 33431, USA}
\begin{abstract}
We show that the standard Hamiltonian of isotropic loop quantum cosmology is selected by physical criteria plus one choice: that it have a `minimal' number of terms. We also show the freedom, and boundedness of energy density, even when this choice is relaxed. A criterion used is covariance under dilations, the continuous diffeomorphisms remaining in this context, which are not canonical but conformally canonical transformations. We propose how to implement conformally canonical transformations in quantum theory. Removal of the infrared regulator yields independence of ordering ambiguities.
\end{abstract}
\maketitle

\section{Introduction}

The epistemic value of `simplicity' in a theory --- in the sense of parsimony of postulates --- goes beyond aesthetics. 
Simplicity is central to the effectiveness of the scientific method itself.
Given a prediction from a theory, there is the question: What happens if the prediction fails? How should the theory be modified? The more postulates in a theory, the more unmanageable this part of the scientific method becomes. 
Cast another way: another epistemic value central to science is that
a theory make `risky predictions' \cite{popper1962}. The fewer the postulates, the fewer ways there are to modify the theory in the face of a 
negative result, and hence the greater the risk.  

The role of \textit{uniqueness theorems} is to reduce a theory to a minimal set of postulates, thus bringing out its simplicity. 
With predictions starting to be made in loop quantum cosmology (LQC)
\cite{agullo2018, abs2017, agullo2015, Agullo:2013ai}, 
it is thus important to have uniqueness theorems for LQC.
Previous works \cite{eht2016, eh2016, ac2012} have addressed the uniqueness of the kinematics of LQC. The present work extends those results to include dynamics.  More specifically, the present work will show that the predominant model of LQC is uniquely determined by basic physical principles, together with only two choices.

Loop quantum gravity is an approach to quantum gravity in which Einstein's fundamental principle of \textit{diffeomorphism covariance} is central. 
Loop quantum cosmology is a quantum theory of the quantum gravitational degrees of freedom at the cosmological level based on the principles of loop quantum gravity. 
The predominant model for loop quantum cosmology is the so-called `improved dynamics', introduced in 2006 \cite{aps2006}. 
With the results of this letter, it is established that both the kinematics and dynamics of this model are uniquely determined by the following physical principles:
\begin{itemize}
\item \textit{(Residual) diffeomorphism covariance} of both the kinematical 
framework as well as the Hamiltonian constraint $\hat{H}$.
\item that $\hat{H}$ should be \textit{Hermitian}.
\item that $\hat{H}$ should have the \textit{correct classical limit}.
\end{itemize}
together with the following two choices:
\begin{enumerate}
\item The holonomy-flux algebra of observables \cite{lost2006, eht2016} should act cyclically in the kinematical quantum theory. This is the only place where loop quantum gravity enters into the assumptions. We call this the \textit{loop hypothesis}.  
\item The number of terms in $\hat{H}$, appropriately defined, should be \textit{minimal}. 
\end{enumerate}
The first of the above two choices, via the kinematical uniqueness theorems
\cite{eht2016, eh2016}, implies that the kinematical Hilbert space of states is that of Bohr's almost periodic functions \cite{abl2003}. In the present work, which focuses on dynamics, this choice thus implies that the Hamiltonian constraint operator should preserve this space of states.

Part of what makes the present uniqueness theorem possible is that, in loop quantum cosmology, one must
take the limit of \textit{large volume} of the fiducial cell, which serves as an infrared cut-off \cite{as2011a}. 
Specifically, the commutator among the basic variables in loop quantum cosmology scales as the inverse
of the volume of the fiducial cell \cite{rw2013, ev2017}, so that in the limit in which the infrared regulator is removed, 
\textit{all operator ordering ambiguities} in the definition of $\hat{H}$ are removed.
This is what allows the present uniqueness result to be stronger than that
in the prior work \cite{ev2017}, where only uniqueness 
up to leading and subleading orders in $\hbar$ was achieved.
Similar reasoning was used in \cite{as2011a}, where the authors point out that inverse volume corrections do not have any physical meaning because they are cell-dependent and vanish once the regulator is removed.

A second key element of the present work is a quantum equation expressing covariance of a given operator with respect to residual diffeomorphisms in LQC.
These residual diffeomorphisms --- dilations --- are not canonical but, rather, \textit{conformally canonical} transformations. 
We introduce a novel method for implementing such transformations in quantum theory which strictly generalizes the standard way of implementing canonical transformations. 
The resulting quantum covariance condition on the Hamiltonian constraint
eliminates the need for certain technical assumptions which were needed in 
the previous work \cite{ev2017} and is much simpler than the condition introduced there, but is otherwise equivalent.

We furthermore note that the question of uniqueness of dynamics in LQC was first investigated much earlier in the work \cite{cs2008a}, which showed how residual diffeomorphism invariance selected the standard dynamics of LQC \cite{aps2006} from a one-parameter family of possible dynamics. The present work starts from no such restriction.  

It is also important to mention that if the single choice being imposed on the dynamics --- minimality --- is removed, then other models also become possible, in particular the `$\overline{\mu}$' versions of the dynamics proposed and investigated in
\cite{ydm2009, adlp2018, lsw2018, lsw2018a, agullo2018, deharo2018, ss2018}.
The present work gives a compact parameterization of the possible
Hamiltonian constraints when minimality is relaxed.
We note that even when minimality is relaxed, the Big Bang singularity 
 is resolved in the sense that energy density is bounded.

\section{Background}
In this section we briefly review the required background material (for more details, see \cite{bojowald2008,as2011a,as2017}). In loop quantum gravity the gravitational phase space variables are given by an $\SU$ connection $A^i_a$ and a densitized triad $E^a_i$. We will consider the simple, $k=0$, spatially homogeneous and isotropic model. Given that the fields are homogeneous on a non-compact slice, the consistent Hamiltonian treatment requires introducing an infrared regulator for integrals. A cubical fiducial cell $\mathcal{V}$ provides such a regulator, which must be removed before extracting physical results from the theory. Let $\fq_{ab}$ be an arbitrarily chosen flat fiducial metric with determinant $\fq$ and $V_o$ the volume of the fiducial cell with respect to this metric. Then by fixing the gauge we can write
\begin{align*}
A^i_a = c V^{-\frac{1}{3}}_o \fom^i_a \, , \quad\quad\quad E^a_i = p V^{-\frac{2}{3}}_o \sqrt{\fq} \fe^a_i \, ,
\end{align*}
where $(\fom^i_a,\fe^a_i)$ are the orthonormal co-triads and triads compatible with $\fq_{ab}$ and adapted to the edges of the cell 
$\mathcal{V}$. Thus, the phase space is two-dimensional and parameterized by $(c,p)$. The non-vanishing Poisson bracket is
\begin{align*}
\{ c, p \} =  \frac{\kappa \gamma}{3} ,
\end{align*}
where $\kappa=8\pi G$ with $G$ the Newton constant and $\gamma$ is the Barbero-Immirzi parameter. Because of the underlying symmetries only the Hamiltonian constraint remains to be imposed. When the lapse function is chosen to be $N=|p|^{3n/2}$ with $n$ a real parameter, the gravitational part of the Hamiltonian constraint is given by
\begin{align*}
H = \frac{-3}{\kappa \BI^2} |p|^{\frac{3n+1}{2}} c^2.
\end{align*}
The group of diffeomorphisms preserving the above gauge-fixing, and 
acting non-trivially on $(c,p)$, is generated by 
parity and the one-parameter family of dilations. 
Parity is defined by 
$\Pi_*\mathring{e}^a_i=-\mathring{e}^a_i$ and 
$\Pi(p_o)=p_o$ with $p_o$ an arbitrary center,
with resulting action $(c,p)\mapsto(-c,-p)$.
The dilations are diffeomorphisms generated by the `radial' vector field $r^a$ defined by
$\partial_br^a=\delta^a_b$ and $r^a(p_o)=0$, where $\partial_a$ is the covariant derivative determined by $\mathring{q}_{ab}$. 
The resulting action is 
$(c,p)\mapsto(e^{-\lambda}c,e^{-2\lambda}p)$ with real parameter 
$\lambda$, with $H$ transforming as 
\begin{align}
\label{eq:dilcov}
H(e^{-\lambda}c,e^{-2\lambda}p) = e^{-3(n+1)\lambda}H.
\end{align}

As shown in \cite{eht2016}, the requirement of diffeomorphism symmetry selects the unique representation of the reduced holonomy-flux algebra used in LQC. This representation is given by the Bohr Hilbert space of almost periodic functions on $\R$. The basic phase space functions with quantum operator analogues are $p$ and $e^{i\mu c}$. The eigenstates $|p\rangle$ of the 
$\hat{p}$ operator form an orthonormal basis of this Hilbert space. The action of the basic operators on these eigenstates is $\hat{p}|p\rangle=p|p\rangle$ and 
$\widehat{e^{i\mu c}}|p\rangle=|p+\frac{\kappa\BI\hbar}{3}\mu\rangle$.

One can extend the definition of the basic operators to include operators of the form $\widehat{e^{if(p)c}}$, such that they map a given momentum eigenstate $|p\rangle$ to $|F(p)\rangle$, where $F(p)$ is the flow, 
evaluated at unit time,
generated by the vector field $\kappa\BI{\hbar}f(p)\frac{d}{dp}$.
In particular, if one chooses $f(p)=\tilde{\mu}|p|^{-1/2}$, one can define 
$\widehat{e^{i\tilde{\mu}b}}$ where $b=|p|^{-1/2}c$, the variable conjugate
to the signed volume $v=\sgn(p)|p|^{3/2}$, 
$\{b,v\}=\frac{1}{2}\kappa\gamma$,
so that the action on $\hat{v}$ eigenstates is 
$\widehat{e^{i\tilde{\mu}b}}|v'\rangle=|v'+\frac{\kappa\gamma\hbar}{2}\tilde{\mu}\rangle$
\cite{aps2006, as2017}. $b/\gamma$ is the Hubble rate and contains all dilation-invariant gravitational information.

\section{Selection of the Quantum Hamiltonian Constraint}
\label{sec:new}

\subsection{Residual diffeomorphism covariance and Hermiticity}
\label{sec:newcov}

Classically, the flow of a phase space function $F$ under the canonical transformation generated by the Hamiltonian vector field $X_{\Lambda}$, associated to a phase space function $\Lambda$, is given by
\begin{align*}
\dot{F} = \mathcal{L}_{X_{\Lambda}} F =  \{ \Lambda, F \}.
\end{align*}
This has the standard quantization
\begin{align*}
\dot{\hat{F}} = \frac{1}{i\hbar} \left[ \hat{\Lambda}, \hat{F} \right] .
\end{align*}

Isotropic dilations are not canonical, but they are \textit{conformally} canonical: they are generated by a vector field of the form
\begin{align*}
X = \omega X_{\Lambda},
\end{align*}
whose corresponding flow thus takes the form
\begin{align*}
\dot{F} = \mathcal{L}_{X} F = \omega \mathcal{L}_{X_{\Lambda}} F = \omega \{ \Lambda, F \}.
\end{align*}
Because, in the case considered in this paper, $\Lambda$ turns out to depend on $c$ so that only its exponential is well-defined in quantum theory, we rewrite the above equation as
\beq
\label{eq:flowexp}
\dot{F} = \omega \frac{1}{i\mu} e^{-i \mu \Lambda} \{ e^{i \mu \Lambda}, F \}.
\eeq
This leads to the quantum equation
\beq
\label{eq:quantflowexp}
\dot{\hat{F}} = \hat{\omega} \star \left(\frac{-1}{\mu \hbar}  
\widehat{e^{-i \mu \Lambda}} \left[
\widehat{e^{i \mu \Lambda}}, \hat{F} \right] \right),
\eeq
where $\star$ denotes a choice of ordering for operator products. Note that equation \eqref{eq:flowexp} is independent of $\mu$; however, the quantization \eqref{eq:quantflowexp} is not. In the case when 
$\omega=\text{const}$ and one uses the Schr\"{o}dinger representation, \eqref{eq:quantflowexp} reduces to the standard flow generated by 
$\Lambda$ only in the $\mu\rightarrow0$ limit, and hence we make this choice:
\beq
\label{eq:quantcond}
\dot{\hat{F}} = \frac{-1}{\hbar} \hat{\omega} \star \lim_{\mu \rightarrow 0} \left( \frac{1}{\mu}  \widehat{e^{-i \mu \Lambda}} \left[ 
\widehat{e^{i \mu \Lambda}}, \hat{F} \right] \right).
\eeq

Let $X$ be the generator of dilations, so that $\mathcal{L}_Xp=-2p$ and 
$\mathcal{L}_Xc=-c$, and write $X=\omega(p,c)X_{\Lambda}$. Because $c$ is not well-defined in quantum theory, we make $\omega$ independent of $c$ which together with the requirement that $X_{\Lambda}$ be a generator of canonical transformation determines $\omega$ up to an overall factor $M$: 
$\omega=-Mv$. Because of the natural appearance of the variable $v$, we will use the $(b,v)$ variables at this point. Then $\Lambda$ is determined up to an additive constant $l$, $\Lambda=\frac{6}{\kappa\BI}(M^{-1}b+l)$, and we get
\begin{align*}
X = -M v X_{\frac{6}{\kappa \BI} (M^{-1} b + l)}.
\end{align*}
The Hamiltonian flows under the action of dilations as 
$\mathcal{L}_XH=-3(n+1)H$, and thus we impose the covariance condition 
$\dot{\hh}=-3(n+1)\hat{H}$. Equation \eqref{eq:quantcond}, for 
$\hat{F}=\hh$, then gives
\begin{align*}
\begin{split}
\frac{-M}{\hbar} \hat{v} \star \lim_{\mu \rightarrow 0} \left(\frac{1}{\mu}  \reallywidehat{e^{-i  \frac{6\mu}{\kappa \BI} (M^{-1} b + l)}} \left[ 
\reallywidehat{e^{i \frac{6 \mu}{\kappa \BI} (M^{-1} b + l)}}, \hh \right] \right)& \\  = 3(n+1) &\hh .
\end{split}
\end{align*}
As expected, $l$ drops out of the equation. Rescaling $\mu$ by 
$\frac{M\kappa\BI}{6}$ we obtain
\begin{align}
\label{eq:gencov}
2 \hat{v} \star \lim_{\mu \rightarrow 0} \left(\frac{1}{\mu}  
\widehat{e^{-i \mu b}} \left[ 
\widehat{e^{i \mu b}}, \hh \right] \right) = - \hbar \kappa \BI (n+1) \hh.
\end{align}
As mentioned above, there is an ordering ambiguity in the product $\star$. For now we choose the Weyl ordering,
$\hat{v}\star\hat{O}:=\frac{1}{2}\left(\hat{v}\hat{O}+\hat{O}\hat{v}\right)$, and address alternative choices in the next section:
\begin{align}
\begin{split}
\label{eq:covquantum}
 - \hbar & \kappa \BI (\!n\!\!+\!\!1\!) \hat{H}\!\! =\! \! \lim_{\mu \rightarrow 0} \!\!\left(\!\hat{v} \widehat{e^{-i \mu b}}\!\! \left[ \widehat{e^{i\mu b}}, \!\hat{H} \right] \!\!+\! \widehat{e^{-i \mu b}}\!\! \left[ \widehat{e^{i\mu b}}, \!\hat{H} \right] \!\!\hat{v}\!\right)\!\!.\!\!\!\text{}
\end{split}
\end{align}

Equation \eqref{eq:covquantum} can be rewritten in terms of the matrix elements of the operator $\hh$ (in the $|v\rangle$ basis),
\begin{align*}
-(n+1) & H(v'',v') = \\
&\frac{v'+v''}{2} \lim_{\tm \rightarrow 0} \frac{1}{\tm} \left(H(v'',v') - H(v''+\tm, v'+\tm)\right),
\end{align*}
where $\tm=\frac{\mu\hbar\kappa\BI}{2}$.  Then, by using the substitution 
$f_w(u)=H(w+u,u)$, we obtain the differential equation,
\begin{align*}
\frac{w+2u}{2} f'_w(u) = (n+1) f_w(u).
\end{align*}
The general solution to this equation is
\begin{align}
\label{eq:Hsol}
H\!(v''\!,\!v')\! = \!f_{v''\!-v'}(v') \!=\! B_{v''\!-v'}(\sgn\! (v''\!\!\!+ \!v')\!)\! \left|\tfrac{v''\!+ v'\!{}}{2}\right|^{n+1}
\end{align}
for some functions $B_w(\sigma)$.

Next we impose that $\hh$ be Hermitian and parity invariant.
Hermiticity implies
\begin{align*}
\overline{H(v',v'')} = H(v'',v').
\end{align*}
Parity invariance implies
\begin{align*}
\bs
\Pi \hh \Pi &= \hh \implies \langle v'' |\Pi \hh \Pi | v' \rangle = \langle v'' |\hh | v' \rangle \\ &\implies H(-v'',-v')=H(v'',v').
\es
\end{align*}
These two conditions together force
$B_w(\sigma)$ to take the form:
\begin{align*}
B_w(\sigma) = a_{|w|} + i\sigma \sgn(w) b_{|w|}
\end{align*}
with $a_{|w|},b_{|w|}$ real.

Finally, we impose that the operator $\hh$ map the $v=0$ eigenstate within the Bohr Hilbert space --- that is, to a countable linear combination of $\hat{v}$ eigenstates with square summable coefficients.  This leads to
\begin{align*}
a_{|w|} &= \tilde{a}_0 \delta_{|w|, 0} + \sum_{i=1}^N \tilde{a}_i \delta_{|w|, v_i} ,\\
b_{|w|} &=  \sum_{i=1}^N \tilde{b}_i \delta_{|w|, v_i}
\end{align*}
with $N$ possibly infinite, $\tilde{a}_0,\tilde{a}_i,\tilde{b}_i$ a 
square summable set of real numbers, and $v_i>0$.

Bringing together the results of the previous paragraphs, we obtain the matrix elements of 
$\hh$,
\begin{align}
\label{eq:matel}
H(v'',&v') = \tilde{a}_0 |v'|^{n+1} \\ 
\nonumber + &\sum_{i=1}^N \left(\tilde{a}_i + i \tilde{b}_i \sgn\left(v'+\frac{v_i}{2}\right)\right) \left\lvert v'+\frac{v_i}{2} \right\rvert^{n+1} \delta_{v'',v'+v_i} \\ 
\nonumber +& \sum_{i=1}^N \left(\tilde{a}_i - i \tilde{b}_i \sgn\left(v'-\frac{v_i}{2}\right)\right) \left\lvert v'-\frac{v_i}{2} \right\rvert^{n+1} \delta_{v'',v'-v_i}.
\end{align}
Therefore, we can write the Hamiltonian $\hh$ as
\begin{align}
\label{eq:selham}
\begin{split}
\hh  = \sum_{i=1}^N
\widehat{e^{i \frac{\tilde{A}_i}{2} b}}\left(\tilde{a}_i + i \tilde{b}_i \sgn(\hat{v})\right)&|\hat{v}|^{n+1} 
\widehat{e^{i \frac{\tilde{A}_i}{2} b}} \! \\
&+ \hc  + \tilde{a}_0 |\hat{v}|^{n+1}
\end{split}
\end{align}
where $\tilde{A}_i=\frac{2v_i}{\hbar\kappa\BI}$ and $\hc$ stands for Hermitian conjugate.

We will now go back to using the standard $(c,p)$ variables to facilitate comparison with APS\cite{aps2006}. 
Notice that in the equation \eqref{eq:selham}, the following quantization prescription naturally appears:
\begin{equation}
\label{eq:qmap}
\reallywidehat{g(p)e^{if(p)c}}:= \widehat{e^{if(p)c/2}} \widehat{g(p)}
\widehat{e^{if(p)c/2}}.
\end{equation}
For brevity, we use this prescription to write expressions for $\hh$ in 
what follows.

We define a ``classical analogue'' of $\hh$ to be
an element of its preimage under a quantization map.
Using the quantization prescription \eqref{eq:qmap}, the classical analogue 
is
\begin{align}
\label{eq:ca}
\rule{0em}{0em} \hspace{-0.3em} H \!\! = \!\! \sum_{i=1}^N \!\left(\!\tilde{a}_i \! + \! i \tilde{b}_i \sgn\!(p)\!\right)\! |p|\rule{-0.3em}{1.05em}^{\!\frac{3(\!n\!+\!1\!)}{2}}\! e^{i\!\tilde{A}_i\! \frac{c}{\sqrt{|p|}}}\! + \cc \!+ \! \tilde{a}_0 |p|\rule{-0.3em}{1.05em}^{\!\frac{3(\!n\!+\!1\!)}{2}}
\hspace{-0.5em}, 
\hspace{-0.2em}\rule{0em}{0em}
\end{align}
where $\cc$ stands for complex conjugate. Note that this $H$ transforms as expected under the action of dilations (see \eqref{eq:dilcov}). We thus conclude that the method used in 
this paper to 
impose that $\hh$ be dilation covariant --- condition \eqref{eq:covquantum} ---
is in fact equivalent to the method used 
to impose such covariance 
in \cite{ev2017}, while at the same time eliminating the need for certain technical assumptions that were required in \cite{ev2017} and leading to a much simpler argument.

\subsection{Single length scale and correct classical limit}

To take the classical limit we let the coefficients 
$\tilde{a}_i,\tilde{b}_i,\tilde{A}_i$ depend on the classicality parameter $\lp:=\sqrt{\hbar G}$: 
$\tilde{a}_i(\lp),\tilde{b}_i(\lp),\tilde{A}_i(\lp)$.
We assume now that $\lp$ is the only length scale in the theory. Dimensional arguments easily lead to $\tilde{A}_i(\lp)=A_i\lp$ and 
$\tilde{a}_0(\lp)=a_0/(G\lp^2),\tilde{a}_i(\lp)=a_i/(G\lp^2),\tilde{b}_i(\lp)=b_i/(G\lp^2)$. This yields
\begin{align}
\label{eq:finalnosimp}
\bs
\hat{H} \!\! = \frac{\lp^{-2}}{G}\! \Bigg(\!\sum_{i=1}^{N} &
\reallywidehat{\left(a_i \!+\! i b_i \sgn\!(p)\!\right) |p|^{\frac{3(n+1)}{2}} \! e^{i \!A\!{}_i \lp \frac{c}{\sqrt{|p|}}}} \\
&\text{}\hspace{2cm}+\hc + a_0 \reallywidehat{|p|^{\frac{3(n+1)}{2}}}\Bigg). 
\es
\end{align}
We next consider a classical analogue of $\hat{H}$. As we are only interested in the classical limit of this analogue ($\lp\rightarrow0$, $\hbar\rightarrow0$), it 
does not matter which ordering is chosen.
Using the ordering 
\eqref{eq:qmap}, one obtains
\begin{align*}
\bs
H \!\!= \frac{\lp^{-2}}{G} \!\Bigg(\!\sum_{i=1}^{N} &\left(a_i \!+\! i b_i\! \sgn\!(p)\!\right)\! |p|^{\frac{3(n+1)}{2}} \! e^{i \!A\!{}_i \lp \frac{c}{\sqrt{|p|}}} \\
&\text{}\hspace{2cm} + \cc + a_0 |p|^{\frac{3(n+1)}{2}}\Bigg).
\es
\end{align*}

We expand the exponentials in powers of $\lp$ and match the classical limit to the classical Hamiltonian $H=\frac{-3}{8\pi G\gamma^2}|p|^{\frac{3n+1}{2}}c^2$:
\begin{widetext}
\begin{align*}
\lim_{\lp \rightarrow 0} \frac{\lp^{-2}}{G} |p|^{\frac{3(n+1)}{2}} \left( a_0 + \sum_{i=1}^{N} \left[ 2a_i - 2b_i  A_i \lp \sgn(p)
\frac{c}{\sqrt{|p|}} - a_i A^2_i \lp^2 \frac{c^2}{|p|} + \BO(\lp^3)\right]\right) = \frac{-3}{8\pi G \gamma^2} |p|^{\frac{3n+1}{2}}c^2.
\end{align*}
\end{widetext}
This gives the following conditions:
\begin{align}
\label{eq:lpneg2cond}
a_0 + \sum_i 2a_i &= 0  \\
\label{eq:lpneg1cond}
\sum_i A_i b_i &= 0 \\
\label{eq:lp0cond}
\sum_i A^2_i a_i  &= \frac{3}{8\pi \gamma^2}.
\end{align}

The above class of Hamiltonians, selected \textit{only} by physical criteria and the loop hypothesis, is the first result of this paper.
Note, in particular, for $N=4$, the `$\overline{\mu}$' versions of the Hamiltonians studied in
\cite{ydm2009, adlp2018, lsw2018, lsw2018a, agullo2018, deharo2018, ss2018} 
are included in our framework,
while `$\mu_o$' versions of Hamiltonians \cite{aps2006b, dl2017} are excluded.

Let us consider a general Hamiltonian in this class. The classical analogue of such Hamiltonian as defined above can be viewed as an effective LQC Hamiltonian $H_{grav}$. Using 
$H_{grav}+H_{matt}=0$ we get
\begin{align*}
\frac{\lp^{-2}}{G} \!\Bigg(\!\sum_{i=1}^{N} &\left(a_i \!+\! i b_i\! \sgn\!(p)\!\right)\! |p|^{\frac{3(n+1)}{2}} \! e^{i \!A\!{}_i \lp \frac{c}{\sqrt{|p|}}} \\
&\text{}\hspace{2cm} + \cc + a_0 |p|^{\frac{3(n+1)}{2}}\Bigg) = H_{matt}.
\end{align*}
For any minimally coupled matter, $H_{matt}$ is related to the matter
energy density $\rho$ by
$H_{matt}=2N|p|^{\frac{3}{2}}\rho$. We obtain
\begin{align*}
\lp^{-2} \!\Bigg(\!\sum_{i=1}^{N} &\left(a_i \!+\! i b_i\! \sgn\!(p)\!\right) \! e^{i \!A\!{}_i \lp \frac{c}{\sqrt{|p|}}} + \cc + a_0 \Bigg) = G\rho,
\end{align*}
For fixed $p$, because the set of coefficients $a_0,a_i,b_i$ are square summable, the left-hand side is an almost periodic function in $|p|^{-1/2}c$,
and so is bounded. Thus, matter density is bounded, so that the Big Bang singularity is resolved in at least this sense. 
Thus, the present work shows that physical principles together with the choice of the holonomy-flux algebra are
\textit{by themselves already} sufficient to ensure one sense of singularity resolution.

\subsection{Minimality}

Now we introduce the second key choice: that the number of terms $N$ be the \textit{smallest} such that the \crefrange{eq:lpneg2cond}{eq:lp0cond} are satisfied. This can be viewed as an implementation of Occam's razor. 
Then $\hat{H}$ is unique up to a single parameter $A$:
\begin{align*}
\hat{H} = \frac{3}{4 \pi A^2 G \BI^2 \lp^{2}} 
\left(\reallywidehat{|p|^{\frac{3(n+1)}{2}}e^{i A\lp \frac{c}{\sqrt{|p|}}}} + \hc -2\reallywidehat{|p|^{\frac{3(n+1)}{2}}}\right).
\end{align*}
In loop quantum gravity the area operator has the minimum eigenvalue $\Delta\lp^2$ with 
$\Delta$ a dimensionless number. If the parameter $A$ is chosen to be $2\sqrt{\Delta}$ and we choose the lapse with $n=0$, we obtain 
\textit{exactly} the `improved dynamics' Hamiltonian introduced in APS \cite{aps2006}.

\section{Ordering ambiguity and the role of the large-volume limit}
In the previous sections we assumed a particular ordering prescription for
the operator product $\star$ (see \eqref{eq:covquantum}). 
We will now address this apparent ambiguity by considering alternative choices. 
Specifically, we demonstrate that in the final quantum theory of cosmology this choice bears no physical significance.

Let us choose an alternative ordering in \eqref{eq:gencov}. A general ordering for the operator product $\hat{v}\star\hat{O}$ for $\hat{O}$ 
arbitrary can be written as
\begin{align*}
\hat{v} \star \hat{O}  =\sum_i \alpha_i \hat{v}^{\lambda_i} \hat{O} \hat{v}^{1-\lambda_i}
\end{align*}
with coefficients $\alpha_i$ such that $\sum_i\alpha_i=1$. Then 
\eqref{eq:gencov}, in terms of the matrix elements of the Hamiltonian with alternative ordering 
$\hh_a$, reads
\begin{align*}
\bs
G(v'', v') \lim_{\tm \rightarrow 0} \frac{1}{\tm} \Big(H_a(v'',v') &- H_a(v''+\tm, v'+\tm)\Big) \\ &= -(n+1) H_a(v'',v'),
\es
\end{align*}
where $G(v'',v')=\sum_i\alpha_i(v'')^{\lambda_i}(v')^{1-\lambda_i}$. By using the substitution $f_w(\eta)=H_a(w(\eta+1)/2,w(\eta-1)/2)$, we get the differential equation
\begin{align*}
K(\eta) f'_w(\eta) = (n+1) f_w(\eta)
\end{align*}
for $K(\eta):=G(\eta+1,\eta-1)=\sum_i\alpha_i(\eta+1)^{\lambda_i}(\eta-1)^{1-\lambda_i}$.
Let $Z\subset\mathbb{R}$ be the set of zeroes of $K(\eta)$. From the fact that $K(\eta)$ is asymptotic to $\eta$ as $\eta\rightarrow\pm\infty$, one can show that $Z$ is bounded. This, together with the fact that, from its form, $K(\eta)$ is analytic in $\eta$, implies that $Z$ is finite.
Let $\eta_1,\eta_2,\dots,\eta_M$ denote the elements of $Z$, in ascending order, and let 
$\eta_0:=-\infty$
and $\eta_{M+1}:=\infty$. 
The general solution to the above differential equation is then
\begin{align*}
H_{\text{}a}(v''\!,\!v')\!=\! \tilde{B}_{v''\!-v'}\!(j(\eta)\!) g(\eta),
\hspace{1em} g(\eta)\!:= \!\exp\!\left(\!\int^{\eta}_{\text{}\!\!\substack{\rule{0cm}{0.4cm} \\ \eta^{\text{}\!o}\!(j(\eta)\!)}} 
\hspace{-0.5cm}\frac{n\!+\!1}{K\!(\eta')} d\eta'\!\right)
\end{align*}
for some $\tilde{B}_w(j)$ and $\eta^o(j)\in(\eta_j, \eta_{j+1})$,
where $j(\eta)$ is defined by $\eta\in(\eta_{j(\eta)},\eta_{j(\eta)+1})$
and \mbox{$\eta:=(v''+v')/(v''-v')$}. 
The requirement that
$\hat{H}_a$ preserve the Bohr Hilbert space forces 
$\tilde{B}_w(j)=\sum_{i=1}^{N_j}\tilde{B}_{ji}\delta_{w,v_{ji}}$
for some $\{N_j\}\subset \mathbb{N}\sqcup\{\infty\}$, 
$\{\tilde{B}_{ji}\}\subset\mathbb{C}$ and $\{v_{ji}\}\subset\mathbb{R}$, so that
\begin{align}
\label{eq:newmatel}
H_a(v'',v') = \sum_{k=0}^{M} \sum_{i=1}^{N_k}
g_{ki}\left(v'+\frac{v_{ki}}{2}\right) \delta_{v'',v'+v_{ki}}
\end{align}
where
$g_{ki}(v)\!:=\!\tilde{B}_{ki}\delta_{j(2v\!/\!v_{ki}\!),k}\,g(2v/v_{ki})$.
Thus the operator and classical analogue are 
\begin{align}
\label{eq:newh}
\rule{0em}{0em}\hspace{-0.3em}\hat{H}_{a} \! =\!\! \sum_{k=0}^{M} \sum_{i=1}^{N_k}
\reallywidehat{g_{ki}(\!v\!) e^{i \tilde{A}_{ki}b}},
\hspace{0.7em}
H_{a} \!=\!\! \sum_{k=0}^{M} \sum_{i=1}^{N_k}
g_{ki}(\!v\!) e^{i \tilde{A}_{ki}b},
\end{align}
with
$\tilde{A}_{ki}\!:=\!\tfrac{2v_{ki}}{\kappa \gamma \hbar}$, and where 
hat and the classical analogue are again defined as in the end of section 
\ref{sec:newcov}.

Now $v$ is the physical volume of the fiducial cell which serves as the infrared regulator. This regulator does not have any physical significance: it has been introduced only to provide a well-defined symplectic structure for quantization and has to be removed --- that is, the limit 
$v\rightarrow\pm\infty$ taken --- as a final and necessary step in defining the quantum theory. Let us begin by looking at the classical analogue \eqref{eq:newh} in this limit.
Let $\epsilon$ be in $(0,1)$. The form of $K(\eta)$ ensures
\bq
\frac{1}{K(\eta)} - \frac{1}{\eta} = \lo\left(\frac{1}{\eta^{1+\epsilon}}\right).
\eq
Therefore, 
\bq
\int^{\eta}_{\eta^o(j_{\sgn(\eta)})} \frac{d\eta'}{K(\eta')} 
- \log\left\lvert\frac{\eta}{\eta^o(j_{\sgn(\eta)})}\right\rvert = \lo\left(\eta^{-\epsilon}\right),
\eq
where we let $j_-:=0$ and $j_+:=M$.
It follows that
\begin{align*}
& 
\lim_{\eta \rightarrow \pm\infty} \frac{\exp\left(\int^{\eta}_{\eta^o(j_\pm)} \frac{(n+1)d\eta'}{K(\eta')}\right)}{\left\lvert\frac{\eta}{\eta^o(j_\pm)}\right\rvert^{n+1}} =
\\ &\exp\!\left(\!(n\!+\!1)\! \lim_{\eta \rightarrow \pm\infty}\! \left(\int^{\eta}_{\eta^o\!(j_\pm)} \!\frac{d\eta'}{K(\eta')} - \log\left\lvert\frac{\eta}{\eta^o(j_\pm)}\right\rvert\right)\right) \!=\! 1,
\end{align*}
whence
\begin{align}
\label{eq:glim}
g_{j_{\pm}i}(v) \sim \tilde{B}_{j_{\pm}i} 
\left\lvert \frac{2v}{\eta^o(j_\pm) v_{j_{\pm}i}}\right\rvert^{n+1}
=:B_i(\pm) |v|^{n+1}
\end{align}
as $v\rightarrow\pm\infty$. Thus,
holding the Hubble rate $b/\gamma$ constant in this limit,
$H_a\sim\sum_{i=1}^{N_{j_{\pm}}}B_i(\pm)|v|^{n+1}e^{i\tilde{A}_{j_{\pm}i}b}$,
which, upon imposing that $\hat{H}_a$ be hermitian and parity invariant, yields the same classical analogue \eqref{eq:ca}, and therefore the same effective dynamics.

We next ask whether the \textit{exact quantum} Hamiltonians are equivalent in this same limit, precisely in the 
sense that 
\begin{align}
\label{eq:largevollim}
\lim_{(|v''|,|v'|)\rightarrow (\infty,\infty)} \frac{H_a(v'',v')+C}{H(v'',v')+C}= 1
%
%
%
\end{align}
where $C$ is any non-zero constant introduced to avoid division by zero.
We will see that also this much stronger condition is true, as long as the number of terms
in \eqref{eq:newh} is \textit{finite}.
Condition \eqref{eq:largevollim} is equivalent to requiring the limit to hold along any path in the 
$(v'',v')$ plane such that $(|v''|,|v'|)\rightarrow(\infty,\infty)$.
For paths on which $w$ approaches infinity, the result follows immediately from \eqref{eq:matel},\eqref{eq:newmatel} and the finiteness of the number of terms.
For paths on which $w$ is bounded, the condition follows 
from \eqref{eq:matel} and the asymptotic form of 
\eqref{eq:newmatel} implied by \eqref{eq:glim} together with parity invariance and hermiticity.

\centerline{---------}

We thank Andrea Dapor for suggesting to add a remark on singularity resolution, and Maximilian Hanusch for helpful presentational remarks on a prior draft. This work was supported in part by NSF grants PHY-1505490 and PHY-1806290.

\bibliographystyle{apsrev4-1}
\bibliography{lqcd-bib}
\end{document}